\documentclass[journal]{IEEEtran}

\usepackage{xcolor,soul,framed} 

\colorlet{shadecolor}{yellow}
\usepackage[pdftex]{graphicx}
\DeclareGraphicsExtensions{.jpeg,.png,.jpg}

\usepackage[cmex10]{amsmath}
\usepackage{array}
\usepackage{mdwmath}
\usepackage{mdwtab}
\usepackage{eqparbox}
\usepackage{url}
\usepackage{booktabs}
\usepackage{tikz}
\usetikzlibrary{positioning}
\usepackage{hyperref}

\tikzset{
  every neuron/.style={
    circle,
    draw,
    minimum size=0.25cm
  },
  neuron missing/.style={
    draw=none, 
    scale=1,
    text height=0.08cm,
    execute at begin node=\color{black}$\vdots$
  },
}

\usepackage{listings}
\usepackage{xcolor}

\lstdefinelanguage{JavaScript}{
  keywords={typeof, new, true, false, catch, function, return, null, catch, switch, var, const, let, async, await, if, in, while, do, else, case, break, from},
  ndkeywords={class, export, boolean, throw, implements, import, this},
  sensitive=false,
  comment=[l]{//},
  morecomment=[s]{/*}{*/},
  morestring=[b]',
  morestring=[b]"
}

\begin{document}
    \title{Federated Learning using Smart Contracts on Blockchain, based on Reward Driven Approach}
  \author{Monik Raj Behera, Sudhir Upadhyay, Suresh Shetty\\
      {\{monik.r.behera,sudhir.x.upadhyay,suresh.shetty\}@jpmorgan.com}

  \thanks{Monik Raj Behera, Onyx Engineering, J.P.Morgan Chase \& Co, India (e-mail: monik.r.behera@jpmorgan.com).}
  \thanks{Sudhir Upadhyay, Onyx Engineering, J.P.Morgan Chase \& Co, India (e-mail: sudhir.x.upadhyay@jpmorgan.com).}
  \thanks{Suresh Shetty, Onyx Engineering, J.P.Morgan Chase \& Co, India (e-mail: suresh.shetty@jpmorgan.com).}}

\markboth{Federated Learning using Smart Contracts on Blockchains, based on Reward Driven Approach
}{}

\maketitle

\begin{abstract}
Over the recent years, Federated machine learning continues to gain interest and momentum where there is a need to draw insights from data while preserving the data provider's privacy. However, one among other existing challenges in the adoption of federated learning has been the lack of fair,  transparent and universally agreed incentivization schemes for rewarding the federated learning contributors. Smart Contracts on a Blockchain network provide transparent, immutable and independently verifiable proofs by all participants of the network. We leverage this open and transparent nature of smart contracts on a blockchain to define incentivization rules for the contributors, which is based on a novel scalar quantity - federated contribution. Such a smart contract based reward-driven model has the potential to revolutionize the federated learning adoption in enterprises. Our contribution is two-fold: first is to show how smart contract based blockchain can be a very natural communication channel for federated learning. Second, leveraging this infrastructure, we can show how an intuitive measure of each agents’ contribution can be built and integrated with the life cycle of the training and reward process.
\end{abstract}

\begin{IEEEkeywords}
Machine Learning, Blockchain, Federated Learning, Fairness in Federated Learning, Systems infrastructure for Federated Learning
\end{IEEEkeywords}

%
\IEEEpeerreviewmaketitle


\section{Introduction}

\IEEEPARstart{T}{he} concept of \textbf{federated machine learning} was introduced around 2016\cite{konevcny2016federated}. It relies on the principle of remote and distributed execution of machine learning algorithm, and the ability to share and aggregate individual models in a secure and anonymous manner. Therefore, it is implicit that federated machine learning would depend on availability of secure communication channels between remote participants to allow distribution of locally trained  individual models.

\textbf{Blockchain} became popular with launch of bitcoin around 2009\cite{nakamoto2012bitcoin}. Blockchain is a form of distributed ledger technology that relies on honest majority members in a network to validate the accuracy of the executed transactions on the network. It accomplishes this by allowing each of its members to execute a piece of turing complete software code (a.k.a smart contract), in an independent fashion without any external influence or interventions. Although the proposed solution could be extended to other Blockchains, this paper  focuses primarily on Ethereum's implementation. Therefore, Blockchains can help the deployment of federated learning by both bringing dataset on a unique, structured, ledger (with potential privacy layers on it), and by guaranteeing the security, accuracy and correctness of the distribution of the model’s parameters. This could be of particular value in compliance and anti-money laundering cases requiring the reconciliation of multiple sensible dataset, and in which the use of fraud and anomaly detection models could improve manual audits and investigations\cite{fortunato2010community}.

However, because of the distributed nature of validation, in current blockchains, unless it implements a privacy layer, all communications between any two nodes is visible to rest of the nodes of the  network.  Preserving privacy of transaction in a blockchain, while still allowing all nodes to participate in consensus process is a difficult problem to solve. This is an active area of research, and includes technologies such as Zero Knowledge Proof (\textit{zKP}). The foundations of \textit{zKP} is based on interactive proofs, as described in previous works\cite{goldreich1991proofs} \cite{bunz2020zether} \cite{bowe2020zexe}.  However, setup of those continues to be  an onerous process, in real world implementation.

This poses a challenge for federated learning, that requires maintaining the privacy of the individual participant's machine learning models (and model gradients too) and anonymity of the model contributor.  The proposed implementation solves this challenge via dynamically generating asymmetric and symmetric keys for each federated learning round ( details to follow in subsequent sections), with a caveat that the aggregation server node is conceptually akin to a Trusted Execution Environment (TEEs), but at a consortium level\cite{cheng2019ekiden}.

However, even with consortium trusted aggregation server implementation, the risk of lack of contribution in the overall federated learning rounds from individual nodes continues to exist.  In addition, the malicious nodes could potentially send a misleading model that could skew the efficacy of the aggregated models.  One of the potential ways to address the above challenges could be to have the aggregation server detect somehow, such behavior and drop those contributors from the collaboration process. Unfortunately, this solution tends to centralizes the solution and lack of transparency in the overall process. 

The current paper proposes a solution to avoid the above potential pitfalls by leveraging unique and transparent smart contracts design on blockchain to reward  honest/active  ( and  penalize malicious/under performing) participants in the learning process, based on computing a novel, scalar quantity - \textbf{federated contribution}. In our proposed solution, smart contract is responsible for reward ( or penalty) specification and distribution (or fees) as well, through immutable federated contribution records on blockchain.

\section{Related work}
\subsection{Federated Learning}
Federated learning is a distributed machine learning setting where the goal is to train a high quality global model, while training is done locally and privately on individual participant (federated learning client). The training is done across large number of clients. After the local models are trained, the improved local model gradients are sent securely over the network to the federated server for federated aggregation\cite{li2018federated}. This paper is focused primarily on horizontal federated learning.

\subsection{Blockchain}
There are several implementations of blockchain - Bitcoin, Ethereum, Hyperledger Fabric etc.  Ethereum\cite{buterin2014ethereum} is a decentralized, open-source blockchain with smart contract functionality. Ether is the native cryptocurrency of the platform. After Bitcoin, it is the second-largest cryptocurrency by market capitalization. In addition to existing public Ethereum mainnet, there exists Ethereum for Enterprises\cite{swan2018blockchain} that offer certain enterprise desired features. Some  of these features include more control on nodes in the network, higher performance,  difference in cost, node permissioning and different privacy implementation.

In general, there are three main types of Blockchains\cite{wood2014ethereum}
\begin{itemize}
\item \textbf{Public} Blockchain: a blockchain that anyone in the world can read, anyone in the world can send transactions to and expect to see them included if they are valid, and anyone in the world can participate in the consensus process.
\item \textbf{Consortium} Blockchain: a consortium blockchain is a blockchain where the consensus process is controlled by a pre-selected set of nodes.
\item \textbf{Fully private} Blockchains: a fully private blockchain is a blockchain where write permissions are kept centralized to one organization. Read permissions may be public or restricted to an arbitrary extent.
\end{itemize}

Some implementations of Blockchain, such as Ethereumsupport both public and Consortium implementations. Onesuch enterprise/consoritum version of Ethereum is Consen-sys \href{https://consensys.net/quorum/}{Quorum} . The proposed solution in this paper applies to the \textbf{Consortium} (Enterprise) blockchain where the identity of individual nodes are known to the Consortium Operator ( a.k.a. Network Operator).

There are few key components of a Blockchain eco-system that are relevant to this paper. Those include namely Smart Contract, Events and Oracles.

\subsection{Smart Contract on Blockchain}
A {\textbf{smart contract}}\cite{bogner2016decentralised} is simply a program that runs on the Ethereum blockchain. It's a collection of code (its functions) and data (its state) that resides at a specific address on the  blockchain. Smart contract are permission-less, i.e. anyone can write a smart contract and deploy it to the network. The smart contract code is visible to all participants on the  network, and any participant can independently execute that code to validate the outcome\cite{bach2018comparative}. Smart Contracts on Ethereum are written in its own programming language called \textbf{Solidity}. \textbf{Events} on Ethereum\cite{EthereumTutorial_2020} are well defined ways of asynchronously exchanging data among the participants of blockchain network. In Solidity, events\cite{SolidityContracts_git} are dispatched as signals, that the smart contracts can fire. \textit{DApps}, which are essentially decentralized applications, or anything connected to Ethereum JSON-RPC API (an interface exposed by blockchain network for connectivity and programmatic interactions with blockchain), can listen to these events and act accordingly. Event can also be indexed, so that the event history can be searched later.

\subsection{ Events on Blockchain }
In Blockchain, when a transaction is mined, smart contracts can emit events and write logs to the blockchain that the frontend can then process. These events can then be used to communicate with a smart contract from application frontend or other subscribing applications. Events are not considered as a state change on Blockchain, hence they consume very less gas price\cite{pierro2019influence}, in comparison to state change transactions on Blockchain.

\subsection{Federated learning and Blockchain}
There is a growing literature on federated learning implementations through blockchain, indicating a sign of the natural complementarity between these two technologies. Since \cite{zhou2020pirate} proposed using blockchain to maintain the global model with community and reach a consensus, a number of papers \cite{kim2019blockchain,majeed2019flchain,bao2019flchain,li2020blockchain} explored this avenue, but mainly using the blockchain as a safe and coherent storage for the global model, and fail to make full use of the potential of smart contracts to both coordinate the learning, and through that compute measurement functions of how each agent is contributing to the global model. 

For that measurement of contribution framework we build on \cite{chen2018machine} proposal to leverage  the  blockchain  to  evaluate updates from nodes, and potentially penalize malicious nodes. 
Shapley values have shown great results in explaining the contributions of individual features in theoritically any machine learning model. 
It is our hope to further bridge these two literatures, to be able to automatically compute different variations of federated learning contributions through blockchain-based smart contract as communication medium in federated learning settings. Our main contribution is to showcase how a natural infrastructure and life cycle could support these, leveraging the cryptographic, distributed computing, and consensus mechanisms within blockchain.

\subsection{RSA and AES algorithm}
Asymmetric key cryptographic algorithm are popular cryptography techniques, which focus on using public-private key pair for encryption. In RSA algorithm\cite{forouzan2015cryptography}, the fundamental idea is to use a computationally impossible, long prime numbers based public key and private key. Public key can be used to encrypt data, where as it can only be decrypted using private key, which is kept privately and securely with the owner. In case of symmetric cryptography, like of AES algorithm\cite{singh2013study}, the same key would be used for encryption and decryption. Thus special care needs to be given to safeguard the symmetric key itself.

\subsection{Measuring contributions in federated learning}
In recent work\cite{9006179}, measurement of contributions towards improvising the global model in federated learning has been described, for both horizontal and vertical class of federated learning. Authors have discussed the approach of using `deletion method' for horizontal federated learning approach, where the change in testing accuracy is considered by various iterations of training, with each iteration, we remove the data points from a single client. This way, we measure the degradation in model performance, and then infer the contributions accordingly by the federated clients. In case of vertical federated learning, Shapley values are computed for each feature. Shapley values gives a strong quantification of each feature's importance, followed by a mathematical approach of inferring contributions of parties in vertical federated learning. There are, however, a multiplicity of ways in which the Shapley value can be implemented, with very disparate results, as shown in \cite{sundararajan2020many}. Our contribution aims to build a mathematical foundation to compute contribution of participants in federated network by focusing on weights differences using Frobenius Norms, instead of training data differences like in Shapley values. 

\section{Blockchain for federated communication}
Earlier works have proved \textit{REST, RPC, gRPC}\cite{ventre2018sdn} as popular choices for communication between federated server and clients. For more secure  communication, various battle-tested approaches include network firewalls, SSL and token based authentication. Though these methods provide a secure medium,  they lack  the transparency in the communication channel itself, which can be well established by using blockchain. This paper proposes use of smart contracts on (\textit{ethereum based}) Blockchain, which provides a decentralized mechanism of communication between peers and the federated server. While smart contracts provide transparency in the implementation, the communications at  blockchain level are secured via asymmetric cryptography techniques for encryption and modification of state on the blockchain network. This also entrusts the emission of events for communication between federated server and contributor nodes with extended level of security and transparency.

Owing to the ``consortium blockchain network'', where individual participants are trusted enterprises and organizations, there is a likelihood of  participants acting as a \textit{bad} actors is low. Further, in this network of contributors, it is quite possible that there could be wide range of variation in contributions from different participants or even extreme scenario of some participants being only a benefitor, and not contributor ( and vice versa). These are few of the possible scenarios in consortium governed federated learning network. To manage and counter these measures through incentives \textit{(both reward and penalty)}, smart contracts provide a transparent and immutable way of maintaining contributions record on the blockchain. Based on the contribution record from blockchain, a given participant would be either rewarded or penalized through on-chain blockchain tokens, or through off-chain mechanism established by consortium.

\subsection{Performance Characteristics of Blockchain}
A well-known constraint of existing blockchain implementations are around performance, usually measured in throughput and latency. Further, current Blockchain implementations also suffer from  Blockchian Trilema ( a coin termed by Vitalik Buterin) - speed, security and decentralization. These constraints are quite true and applicable for high volume, low latency  networks where speed of transaction is quite critical. However,  sharing models over blockchain in an enterprise network does not require significant high throughput sine the frequency of model updates is expected to be relatively low. In addition, latency is also not a critical factor since aggregation of models from each participant are not necessarily time sensitive. For example, if a model update was missed by the aggregation server in one aggregation cycle, it will be picked up in the subsequent one without losing its impact.

\subsection{Life-cycle of a federated aggregation event}
Federated learning in a consortium network comprises of a consortium trusted and security hardened aggregation server. All the participants, who are clients in the federated learning ecosystem listen to the blockchain events from a smart contract (\textit{ethereum smart contracts, as ethereum is the blockchain network being used}). Table \ref{life-cycle} gives an overview of a typical federated learning round. In the proposed implementation, the Aggregator Server \textbf{manages} the  events that orchestrate  the Federated Learning cycle between clients and aggregator.

Since the orchestration of events is delegated to the aggregator, this design allows individual participants remain lightweight and only focus on local model improvisation. The set of possible events from Aggregator server includes initial distribution of base model, subsequent federated learning cycle events, contribution announcement on the chain and contribution fees notification. 

\begin{table}[t]
\caption{Life-cycle of a federated learning communication event. Time column represents relative timestamp, where increment in timestamp suggests relative increment, not the absolute increment. Event is the respective event type on communication round. `Action On' column shows the expected action for which role, whether client or server. `Publish' column depicts whether the event is published as broadcast on blockchain or not.}
\label{life-cycle}
\vskip 0.15in
\begin{center}
\begin{small}
\begin{sc}
\begin{tabular}{|p{0.1\linewidth}|p{0.4\linewidth}|p{0.15\linewidth}|p{0.15\linewidth}|}
\toprule
Time & Event & Action on- & Publish? \\
\midrule
$t_{k}$ & Initiate FL round & Server & $\surd$ \\
$t_{k+1}$ & Receive Initiate Event & Client(s) & $\times$ \\
$t_{k+2}$ & Encrypt local model; with key received in previous step & Client(s) & $\times$ \\
$t_{k+3}$ & Publish encrypted local model & Client(s) & $\surd$ \\
$t_{k+4}$ & Aggregate encrypted local models & Server & $\times$ \\
$t_{k+5}$ & Broadcast new global model & Server & $\surd$ \\
\bottomrule
\end{tabular}
\end{sc}
\end{small}
\end{center}
\vskip -0.1in
\end{table}

\subsection{Encryption of event data}
In the previous sections, we discussed about how blockchain ensures required security and privacy at its core. Though data on blockchain is immutable, the privacy is not guaranteed, because of the very nature of how blockchain works. A consortium network with enterprise participants, there is a high possibility of participants not comfortable with revealing their model weight gradients to peer participants. If local model weights (\textit{and gradients}) are revealed, this posses risk of revealing statistical properties of the data from individual participant, if not the actual data. 

In order to privately send model weights from participant to federated server on blockchain, in a consortium network, asymmetric encryption using \textbf{RSA} cryptography is proposed in the paper. Each new federated learning round is published as blockchain event, federated aggregation server generates a new set of RSA key pair. The private key of the pair stays with the server, as this will be used by the server in later stages to decrypt the encrypted event message cipher. The public key is sent across the network to all the participants. Participant will generate an AES key for encrypting local train model, and then use the public key received from federated server to encrypt the AES key\cite{khanezaei2014framework}. With RSA keys revolving for every new federated learning round, this decreases the chances of compromising model information over the blockchain, for any given participant on the network. One thing to note here is - only the the event data containing participant's local model weights needs to be encrypted. Other event data like global model, which is sent from server to clients, initiation of federated round broadcast does not require encryption, as it does not contain sensitive data among the peers in blockchain network. 

\subsection{Smart Contract for communication and contribution towards Federated Learning}
In the previous sections, smart contracts have been described as a way of executing Turing complete programs on blockchains. Further, as referred in related work, Ethereum smart contracts also have a concept of events, which are generated after a transaction is mined on the Blockchain network. Events can carry additional information and parameters that can be used by subscribing applications.  It is important to understand that events generated, when published by the participants or the federated server, would be a broadcast. Dynamically generated encryption keys, explained in earlier section would safeguard the privacy interest of the federated clients. Federated clients would have to listen to the agreed upon Ethereum \textit{event} from a smart contract, in order to receive the event and take necessary actions, essentially making smart contracts a primary \textit{pub-sub}\cite{o2007toward} channel of communication.

In the current paper, smart contracts are proposed as a mechanism to maintain transparent, immutable records of \textbf{contributions}, which improvises (\textit{or degrades}) the global federated learning model. The computation to determine the contribution of individual in the federated learning round (\textit{described in subsequent section}) is performed off the blockchain, considering the resource and computation constraints. After the computation is completed, the records are published on blockchain, \textbf{anonymously}, for each participant to consume. This makes the contributions (and indirectly participants expense fees) transparent. The transparency of  contributions from each participant promotes an honest behavior on the network. 

\section{Federated aggregation and contribution}
Federated learning with centralized aggregation server implements various ways to implement federated aggregation\cite{bonawitz2019towards}, like \textit{FedAvg}, \textit{FedSGD}, etc. Assuming the number of clients is quite large (\textit{in order of $10^{5}$ }), if few of the clients are acting as \textit{bad} actors, or clients with noise in training data, their impact towards the global model is smoothed by averaging algorithm. But in case of consortium setting, where number of clients is not that huge (\textit{in order of $10^{2}$ }), the local weights of federated client with noisy data or malicious intention would impact the overall weight(s) of the federated learning model. In order to tackle the challenge, we are proposing a novel way of computing a novel scalar quantity, \textbf{federated contribution} across the network, and using smart contract to publish and store on the blockchain (as discussed in earlier section). As compared to earlier work\cite{9006179}, federated contribution establishes a way to define contribution of each participant in federated learning setting, whether they might have train data of similar statistical properties, or non overlapping (\textit{orthogonal in feature space}) statistical properties.

\subsection{Federated aggregation}
\textbf{FedAvg}\cite{li2019convergence} is one of the popular algorithms in federated aggregation domain, which ensures complete and optimal solution, provided the learning rate and local learning contribution is accurately considered. Inspired from the previous works, our problem formulation for non-IID (\textit{or IID}) data, which assumes more real world problem setup, in a consortium blockchain network, we have formulated our problem as described below - 
\[
    \min_{\theta \in \mathbf{R^{d}}} f(\theta) := \sum_{k=1}^{K}p_{k}F_{k}(\theta)
\]
where $k$ as index for client, $\theta$ as the set of model weight parameters for any generic machine learning algorithm, which can be modified for federated learning setting. $f(\theta)$ is the global objective function, $F_{k}(\theta)$ is the local objective function for client $k$, $p_{k}$ is the learning importance factor - a scalar value to determine individual client's, local model's relative importance, and $K$ is the total number of clients. Considering $p_{k}$ as learning importance factor in federated learning, it is assumed that - 

\[
    \sum_{k=1}^{K}p_{k} = 1
\]

With \textit{fedAvg} as algorithm for aggregation, the individual weight updates for $t+1$ iteration, $l^{th}$ layer can be defined as - 

\[
    w_{t+1, l} \leftarrow w_{t, l} + \alpha\sum_{k=1}^{K}p_{k}\nabla F_{k}(\theta)
\]
where $w_{t,l}$ is weight for $t^{th}$ iteration at $l^{th}$ layer and $\alpha$ is the learning rate parameter.

\subsection{Federated learning contribution}
In previous sections, motivation for \textbf{federated contribution} is discussed. Intuitively, federated computation is a scalar quantity, which depicts the deviation, or divergence of two machine learning models. We will now define the federated contribution mathematically.

\[
    \gamma^{k} := \|\beta^{k}\|_{2}
\]

\[  
\beta^{k} := \langle \|\delta_{1}^{k}\|_{\mathbf{F}}, \quad \|\delta_{2}^{k}\|_{\mathbf{F}}, \quad \ldots \quad, \|\delta_{L}^{k}\|_{\mathbf{F}} \rangle 
\]

\[
    \delta_{l}^{k} := \mathbf{w_{l, t}^{global}} - \mathbf{w_{l, t+1}^{k}}
\]

\[
    \gamma^{k}_{rel} := \frac{\gamma^{k}}{\sum_{k=1}^{K}\gamma^{k}}
\]

where $\gamma^{k}$ is the absolute federated contribution of client $k$, $\|.\|_{p}$ represents $p^{th}$ norm, $\|.\|_{F}$ represents Frobenius norm\cite{dhillon2005generalized}, $L$ represents the final layer's weight matrix of a generic machine learning model. $\delta_{l}^{k}$ represents difference of model weight parameter matrix for $l^{th}$ layer of $k^{th}$ client. $\mathbf{w_{l, t+1}^{k}}$ represents model weight for $l^{th}$ layer of $k^{th}$ client at $t+1^{th}$ iteration. $\mathbf{w_{l, t}^{global}}$ is the model weight for $l^{th}$ layer of global model at $t^{th}$ iteration. $\gamma^{k}_{rel}$ is relative federated contribution of client $k$.

For any generic machine learning algorithm, like linear regression, logistic regression, neural networks, etc, the primary intent is to find a weight matrix, or a set of weight matrices. These weight matrices are computed using loss functions and gradient descent approaches, in various formats depending upon the algorithm. Assuming the representational vector, defined $\beta^{k}$, with minimum size of the set as $1$, we can define an each element of $\beta^{k}$ as Frobenius norm of $\delta_{l}^{k}$. Here, $\delta_{l}^{k}$, as defined previously, represents the element wise difference of weight matrices (order of subtraction doesn't matter, as Frobenius norm computes square of the element). We have considered the element wise difference of both the weight matrices for each layer, as it is inverse operation to the gradient descent and weight updates, where we perform addition of improvements, as described in problem formulation earlier. Frobenius norm is well established method to find the magnitude of a matrix from origin of a hyperspace $\mathrm{R^{d}}$. When we compute Frobenius norm for $\delta_{l}^{k}$, it essentially represents the magnitude of deviation between two weight matrices(global weight matrix of previous iteration and local weight matrix of current iteration, for layer \textit{l}). The vector $\beta^{k}$ represents a set of magnitudes of deviation of local model's weight matrices from global model's weight matrices respectively, as an independent and individual axis. Finally, in order to calculate value $\gamma^{k}$ for client \textit{k}, we calculate \textit{2-Norm}, which is the \textit{euclidean norm} of $\beta^{k}$. This represents the magnitude of distance from origin, which quantifies the combined deviation of local model (\textit{set of weights}) from global model (\textit{again, set of weights}) as a scalar quantity.

Inferring $\gamma^{k}$, in case of federated contribution, federated aggregation server calculates $\gamma^{k}$ for each $k^{th}$ client. If the federated contribution value is relatively high, that means the given client has contributed to a \textit{higher degree}. While describing contribution to a higher (\textit{or lower}) degree in federated contribution, it practically quantifies the contribution in modifying the global model, by training on larger data points, or by training over data points which are having distant statistical properties from earlier training data, or may be higher noise in data. This intuitively can be thought as - the divergence of local model after training on new data points will be more, if higher gradient descent updates are performed. This can be because of variance in new data, use of better data points for local training rounds or greater training size. In case of divergence being relatively smaller, one can infer about participant using noisy or unrelated data points, which may not be acceptable for global federated model. Our framework would thus point at a possible way to fix a key limitation of the Shapley value framework (which we hope to build the link with in a future paper) - the fact that it only provides valuations for points within a fixed data set, and does not account for statistical aspects of the data and does not give a way to reason about points outside the data set.

\section{Experiments}
\begin{figure}[ht]
\vskip 0.2in
\begin{center}
\centerline{\includegraphics[width=\columnwidth]{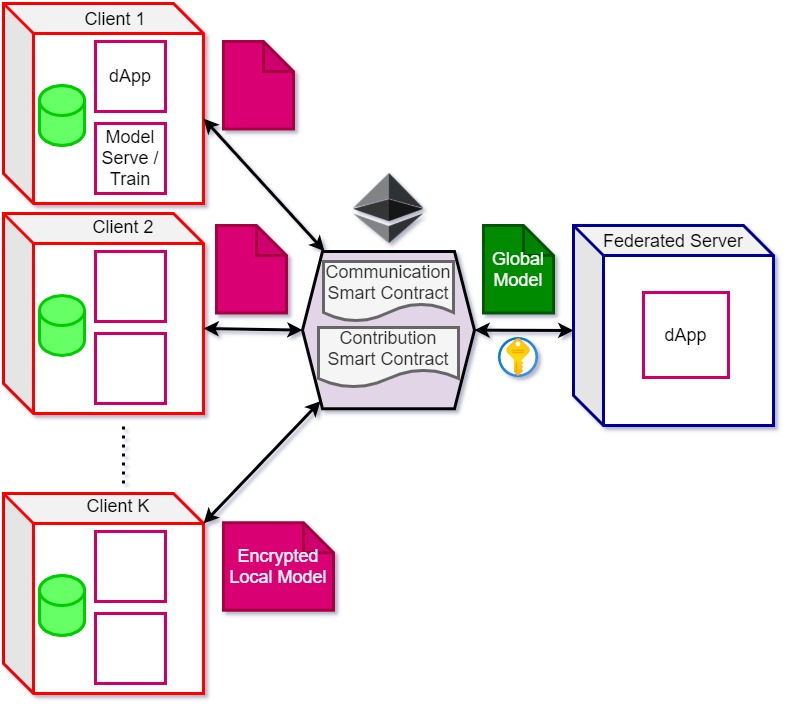}}
\caption{In the above architecture diagram, individual nodes are running decentralized applications, with processes to serve and re-train data models. Two smart contracts, one for communication and one for contribution is deployed on the blockchain. Federated server is responsible to send the dynamically generated RSA key pair's public key, which would be used to encrypt the newly generated AES key of participant, used for model encryption. Encrypted models are published on the blockchain. Even though all the participants can receive the private encrypted model gradients, only the federated aggregation server can decipher and use the model, as it posses the RSA private key of respective public key.}
\label{overview}
\end{center}
\vskip -0.2in
\end{figure}

In the paper, the experiments to validate the hypothesis of using blockchain smart contracts and federated contribution has been carried out with setting up a $(5+1=)6$ ethereum blockchain permissioned nodes setup in AWS cloud. Within this network, one of the node acts as federated aggregation server, and other three nodes act as federated clients. It is a consortium blockchain setup, with each individual client nodes own set of data. Figure \ref{overview} shows the architecture representation of the proposed experimental setup. The federated server node runs a python daemon process which listens to the events generated by smart contracts on the network. Upon receipt of events and based on event types, it either sends the global model to every node as a broadcast, or computes the aggregated version of the global model with latest gradients of local training, from individual federated clients. On  federated clients, a similar python application is being executed, which listens to events and sends the encrypted local model, as required. It also acts as a service, which serves (\textit{exposing remote REST based endpoint, which can be consumed by any other program to predict, based on input data}) machine learning model, and also is responsible for re-training of new batch of data points. The applications, which are responsible for directly interacting with blockchain nodes are termed as ``dApp'', as depicted in figure \ref{overview}, which essentially means decentralized applications. 

Federated  Aggregator server node deploys both s `\textit{\textbf{communication}}' and `\textbf{contribution} \textbf{records}' smart contracts. . All of the federated client nodes leverage the address of these two public contracts as a communication channel both for publishing events and listening to any generated events as well. The interface of the smart contract has been discussed in further.

\subsection{Data description}
In the paper, to test the hypothesis against standard data set, MNIST\cite{lecun-mnisthandwrittendigit-2010} and Fashion-MNIST\cite{xiao2017fashionmnist} data sets have been used. In both the data sets, it has $10000$ test data points, $60000$ train data points and $10$ target classes. We split the training data as $10000$ for initial, ``genesis'' global model and remaining $50000$ as training data for federated clients. For both the data sets, we have performed our experiments in two ways, by splitting our data sets described below - 

\begin{enumerate}
    \vskip -0.2in
    \item Independent and Identical Data: The data is randomly split and distributed across all the federated clients, each having $10000$ training data points with all the $10$ target classes. 
    \vskip -0.2in
    \item Non-Independent and Identical Data: The data is split and distributed across all the federated clients, each of the clients having training data with only $2$ target class.
    \vskip -0.2in
\end{enumerate}

\subsection{Blockchain smart contract interface}
As discussed earlier, we have used two ethereum based smart contracts, one for communication and one for recording contributions transparently on blockchain. Interface for both of the smart contracts have been defined as follows -

\begin{lstlisting}[language=JavaScript,numbers=none]
// Communication : Interface
pragma solidity >=0.8.0 <0.9.0;
contract Communication {
  event BCEvent(
    uint256 timestamp, 
    bool is_encrypted, 
    bytes event_type, 
    bytes body
  );
  function publish(
    uint256 timestamp, 
    bool is_encrypted, 
    bytes memory event_type, 
    bytes memory body
  ) public returns(uint ack) { }
}
\end{lstlisting}

\begin{lstlisting}[language=JavaScript,numbers=none]
// Contribution : Interface
pragma solidity >=0.8.0 <0.9.0;
contract Contribution {
  uint len = 5; //5 federated clients
  uint[] memory _clients = new uint[](5);
  function set_contribution(
    uint client_id, 
    uint relative_contribution
  ) public returns(uint ack) { 
  //only owner(federated server)
  //modifies state of _clients 
  }
  function get_contributions() 
  public view returns (uint memory) { } }
\end{lstlisting}

\subsection{Neural network for image classification}
\begin{figure}
    \centering
    \def\layersep{1.25cm}
    \begin{tikzpicture}[shorten >=1pt,->,draw=black!50, node distance=\layersep]
        \tikzstyle{every pin edge}=[<-,shorten <=1pt]
        \tikzstyle{neuron}=[circle,fill=black!25,minimum size=17pt,inner sep=0pt]
        \tikzstyle{input neuron}=[neuron, fill=green!50];
        \tikzstyle{output neuron}=[neuron, fill=red!50];
        \tikzstyle{hidden neuron}=[neuron, fill=blue!50];
        \tikzstyle{annot} = [text width=4em, text centered]
    
        \foreach \name / \y in {1,...,4}
            \node[input neuron] (I-\name) at (0,-\y) {};
    
        \foreach \name / \y in {1,...,5}
            \path[yshift=0.5cm]
                node[hidden neuron] (H1-\name) at (\layersep,-\y cm) {};
                
        \foreach \name / \y in {1,...,5}
            \path[yshift=0.5cm]
                node[hidden neuron] (H2-\name) at ({2.5cm},-\y cm) {};
                
        \foreach \name / \y in {1,...,5}
            \path[yshift=0.5cm]
                node[hidden neuron] (H3-\name) at ({3.75cm},-\y cm) {};
    
        \node[output neuron,pin={[pin edge={->}]right:Output}, right of=H3-3] (O) {};
    
        \foreach \source in {1,...,4}
            \foreach \dest in {1,...,5}
                \path (I-\source) edge (H1-\dest);
                
        \foreach \source in {1,...,5}
            \foreach \dest in {1,...,5}
                \path (H1-\source) edge (H2-\dest);
        
        \foreach \source in {1,...,5}
            \foreach \dest in {1,...,5}
                \path (H2-\source) edge (H3-\dest);
    
        \foreach \source in {1,...,5}
            \path (H3-\source) edge (O);
    
        \node[annot,above of=H1-1, node distance=1cm] {(1024)};
        \node[annot,above of=H2-1, node distance=1cm] {(512)};
        \node[annot,above of=H3-1, node distance=1cm] {(128)};
        \node[annot,above of=I-1] {(728)};
        \node[annot,above of=O] {(10)};
    \end{tikzpicture}
    \caption{In the architecture, the input layer contains $728$ neurons, after flattening the image. Output layers contains $10$ neurons, as the data set has $10$ image labels or target classes. It contains $3$ fully connected, hidden layers, each having $1024$, $512$ and $128$ neurons respectively. The hidden layers used \textit{relu}, and output layer used \textit{softmax} as the activation functions.}
    \label{nndesign}
\end{figure}
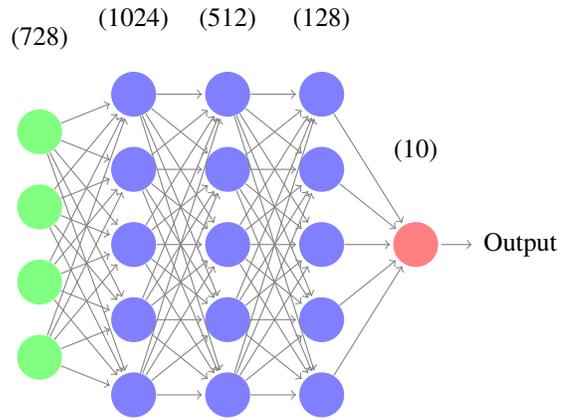

Since the each image data in MNIST and Fashion-MNIST is $28 \mathbf{x} 28$ gray scale image, we have used artificial neural network\cite{deng2013new} for image classification. Figure \ref{nndesign} describes the architecture of the neural network in depth. We have used \textbf{adam}\cite{kingma2014adam} optimizer and \textbf{sparse categorical cross entropy}\cite{louizos2017learning} as loss function, with $5$ training epochs.

\subsection{Results}
Based on the data, blockchain network with smart contracts and neural network architecture described in earlier sections, we performed various experiments to test our hypothesis of computing federated contribution against various settings like higher contribution of a smaller group of participants, lower contributions by a subset of participants, noise in local data, while training by individual nodes. We have also calculated the overall accuracy of image classification of the aggregated model. In our experiments, we have changed the size of training samples, as it is the \textit{variable} of the experiment. For each set of conditional environment for experiments, we have considered results for $4$ different cases. The $4$ cases are as \textit{(i) MNIST data with non-IID split (ii) MNIST data with IID split (iii) Fashion-MNIST data with non-IID split and (iv) Fashion-MNIST data with IID split}. A well known and common observation across all the experimental setting is increase in machine learning model's accuracy performance, with increase number of training samples.  

\begin{figure}[ht]
\vskip 0.2in
\begin{center}
\centerline{\includegraphics[width=\columnwidth]{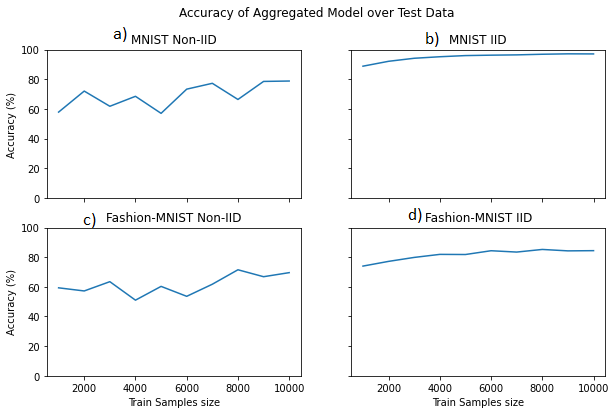}}
\caption{(a) Accuracy of aggregated model in \% vs training data sample size for MNIST data, split in Non-IID format (b) Accuracy of aggregated model in \% vs training data sample size for MNIST data, split in IID format (c) Accuracy of aggregated model in \% vs training data sample size for Fashion-MNIST data, split in Non-IID format (d) Accuracy of aggregated model in \% vs training data sample size for Fashion-MNIST data, split in IID format }
\label{acc}
\end{center}
\vskip -0.2in
\end{figure}

\begin{figure}[ht!]
\vskip 0.2in
\begin{center}
\centerline{\includegraphics[width=\columnwidth]{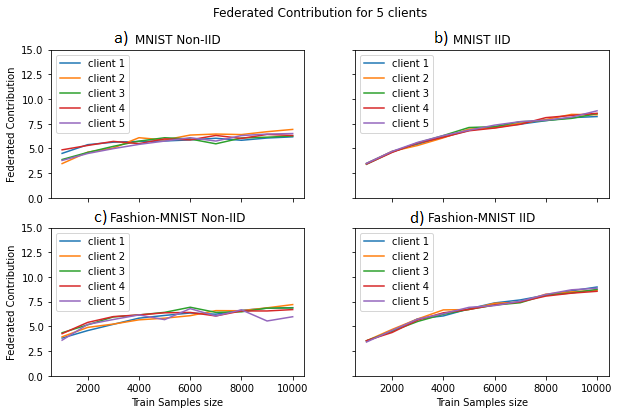}}
\caption{(a) Federated contribution of equally contributing clients vs training data sample size for MNIST data, split in Non-IID format (b)Federated contribution of equally contributing clients vs training data sample size for MNIST data, split in IID format (c) Federated contribution of equally contributing clients vs training data sample size for Fashion-MNIST data, split in Non-IID format (d) Federated contribution of equally contributing clients vs training data sample size for Fashion-MNIST data, split in IID format}
\label{5g0b}
\end{center}
\vskip -0.2in
\end{figure}

Figure \ref{acc} shows the performance of the aggregated machine learning model, from various clients. We do observe better and stable performance in case of data being split, across all the clients in \textit{IID} manner. Figure \ref{5g0b} shows the federated contribution being increasing with increase in training data sample size. Also, since all the clients in this setting have contributed equally, their federated contribution values are quite close. One thing to notice, which is evident across all experiments (\textit{can be observed in the graphs}) is, increase in federated contribution value, with increase in training size. This essentially validates our hypothesis of federated contribution being dependent on number of weight updates, which is directly proportional to higher training data (or training iterations). For distinct visuals, we have shown \textit{client 3} in always green colour, in all the visualizations.  

\begin{figure}[ht!]
\vskip 0.2in
\begin{center}
\centerline{\includegraphics[width=\columnwidth]{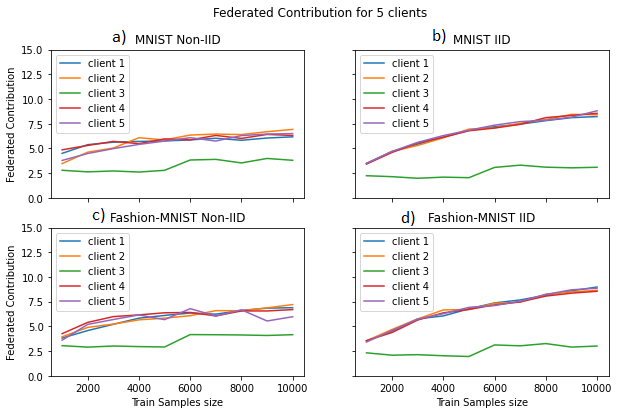}}
\caption{(a) Federated contribution of highly contributing clients, but client 3 vs training data sample size for MNIST data, split in Non-IID format (b)Federated contribution of highly contributing clients, but client 3 vs training data sample size for MNIST data, split in IID format (c) Federated contribution of highly contributing clients, but client 3 vs training data sample size for Fashion-MNIST data, split in Non-IID format (d) Federated contribution of highly contributing clients, but client 3 vs training data sample size for Fashion-MNIST data, split in IID format}
\label{4g1b}
\end{center}
\vskip -0.2in
\end{figure}

\begin{figure}[ht!]
\vskip 0.2in
\begin{center}
\centerline{\includegraphics[width=\columnwidth]{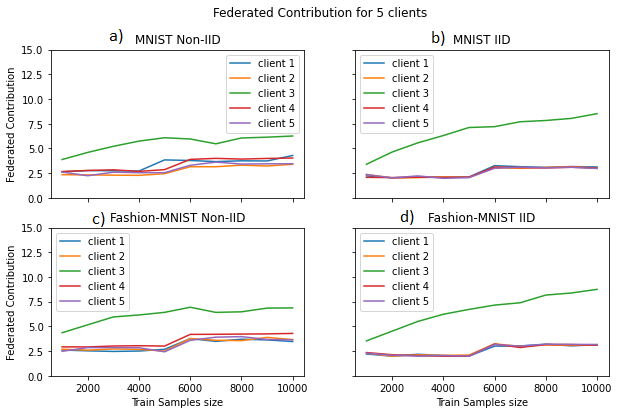}}
\caption{Federated contribution of lowly contributing clients, but client 3 vs training data sample size for MNIST data, split in Non-IID format (b)Federated contribution of lowly contributing clients, but client 3 vs training data sample size for MNIST data, split in IID format (c) Federated contribution of lowly contributing clients, but client 3 vs training data sample size for Fashion-MNIST data, split in Non-IID format (d) Federated contribution of lowly contributing clients, but client 3 vs training data sample size for Fashion-MNIST data, split in IID format}
\label{1g4b}
\end{center}
\vskip -0.2in
\end{figure}

\begin{figure}[ht!]
\vskip 0.2in
\begin{center}
\centerline{\includegraphics[width=\columnwidth]{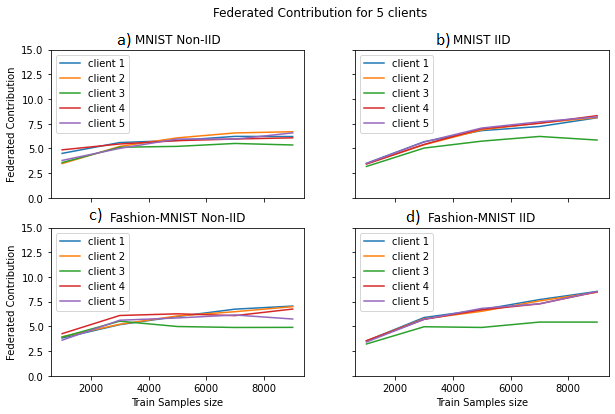}}
\caption{(a) Federated contribution of highly contributing clients, but client 3 with noise vs training data sample size for MNIST data, split in Non-IID format (b)Federated contribution of highly contributing clients, but client 3 with noise vs training data sample size for MNIST data, split in IID format (c) Federated contribution of highly contributing clients, but client 3 with noise vs training data sample size for Fashion-MNIST data, split in Non-IID format (d) Federated contribution of highly contributing clients, but client 3 with noise vs training data sample size for Fashion-MNIST data, split in IID format}
\label{4g1bn}
\end{center}
\vskip -0.2in
\end{figure}

Figure \ref{4g1b} depicts the experiment, where \textit{client 3} is under-performing by training on only $10\%$ of what other participants are training. We can observe how the federated contribution value for \textit{client 3} is very low, as compared to other participants. Figure \ref{1g4b} whereas shows the opposite, where \textit{client 3} trains on $10$ times more data points, puts in extra effort, logically, than other participants. This shows higher value in federated contribution for \textit{client 3}, as relatively compared to others. Figure \ref{4g1bn} is the final setting, where we added Gaussian noise, with noise value between $0$ and $1$, in training data of \textit{client 3}, which essentially shows depletion in federated contribution of \textit{client 3}, relative to others. 

\section{Conclusion}
In our paper, we studied our proposal of using  smart contracts to establish a fair, transparent, secure and immutable incentivization mechanism in a consortium blockchain network for federated learning. We proposed a novel approach to calculate a unique scalar quantity, \textit{federated contribution}, which quantifies the contribution of each participant in federated learning. Federated contribution is compatible with the machine learning algorithms which relies on weight parameters computed by gradient descents. We justified our proposed approach both empirically and theoretically. For future work in the given area, one can extend the federated contribution to non-gradient descents algorithms, or to heterogeneous federated learning. In the proposed method of calculating federated contribution, and using the relative federated contribution values for reward (or penalization) mechanism, we validated that it effectively penalizes under-performing participants, rewards over-performing participants and penalizing participants with noisy or malicious data points. This justifies our proposal of considering federated contribution as an adequate mechanism of quantifying participants' contribution in the  consortium blockchain network. Future work will aim at building further the conceptual bridge between our weight-based contribution measure and Shapley values, under modified axioms that reflect the specificities of federated machine learning settings.

\section*{Acknowledgment}

We would like to thank our colleagues Robert Otter, Nicolas X Zhang, Jiao Y Chang, Tulasi Das Movva, Thomas Eapen for reviews and discussions during the experimentation and evaluation. We would also like to thank J.P.Morgan Chase for supporting us in carrying out this effort. 

\section*{Disclaimer}
This paper was prepared for information purposes by the Onyx Engineering of JPMorgan Chase \& Co and its affiliates (J.P. Morgan), and is not a product of the Research Department of J.P. Morgan. J.P. Morgan makes no explicit or implied representation and warranty and accepts no liability, for the completeness, accuracy or reliability of information, or the legal, compliance, financial, tax or accounting effects of matters contained herein. This document is not intended as investment research or investment advice, or a recommendation, offer or solicitation for the purchase or sale of any security, financial instrument, financial product or service, or to be used in any way for evaluating the merits of participating in any transaction.


%





\ifCLASSOPTIONcaptionsoff
  \newpage
\fi



\bibliographystyle{IEEEtran}
\bibliography{Main}


\vfill


\end{document}